# Near-optimal RNA-Seq quantification

Nicolas L. Bray[1], Harold Pimentel[2], Páll Melsted[3] and Lior Pachter*,[2,4]

1. Innovative Genomics Initiative, UC Berkeley
2. Department of Computer Science, UC Berkeley
3. Faculty of Industrial Engineering, Mechanical Engineering and Computer Science, University of Iceland
4. Departments of Mathematics and Molecular & Cell Biology, UC Berkeley

* Corresponding author: lpachter@math.berkeley.edu

**We present a novel approach to RNA-Seq quantification that is near optimal in speed and accuracy. Software implementing the approach, called kallisto, can be used to analyze 30 million unaligned paired-end RNA-Seq reads in less than 5 minutes on a standard laptop computer while providing results as accurate as those of the best existing tools. This removes a major computational bottleneck in RNA-Seq analysis.**

The first two steps in typical RNA-Seq processing workflows are alignment (to a transcriptome or a reference genome) and estimation of transcript abundances. These steps are time consuming; for example aligning 20 samples, each with 30 million RNA-Seq reads with the program TopHat2[1] takes 23 core hours, while quantification with the companion program Cufflinks[2] takes an additional 10 core hours. Such running times have become impractical for current projects that involve sequence from thousands of samples. While the quantification of aligned reads can be sped up with streaming algorithms[3] or by naïve counting of reads[4], the time consuming alignment step has been indispensible. To circumvent the alignment step, it has recently been proposed to quantify samples by exact matching of read *k*-mers using a hash table[5]. Unfortunately, shredding reads into *k*-mers discards valuable information present in complete reads resulting in a substantial loss of accuracy (Supp. Fig 1).

While the direct use of *k*-mers is inadequate for accurate quantification, the speed of hashing provides hope for much faster, yet accurate, RNA-Seq processing. We therefore asked whether information from *k*-mer within a read could be combined efficiently in a manner that would maintain the accuracy of alignment-based quantification. To address this question, we examined the central difficulty and key requirement for accurate quantification, which is the assignment of reads that cannot be uniquely aligned[6]. Typically, these multi-mapping reads are accounted for using a statistical model of RNA-Seq[6] which probabilistically assigns such reads while inferring maximum likelihood estimates of transcript abundances. However it has been observed that the sufficient statistics for the simplest such models are the compatibilities of reads with transcripts[7]. That is, the necessary information is not *where* inside transcripts the reads may have originated from, but only *which* transcripts could have generated them. This led us to formulate the concept of *pseudoalignment* of reads and fragments.

A pseudoalignment of a read to a set of transcripts $T$ is therefore a subset $S \subseteq T$ without specific coordinates mapping each base in the read to specific positions in each of the transcripts $S$ as would be present in a read alignment. Highly accurate pseudoalignments for reads to a transcriptome can be obtained efficiently using fast hashing of *k*-mers together with the *transcriptome de Bruijn* graph. De Bruijn graphs have been crucial for DNA and RNA assembly[8], where they are usually constructed from reads. However, in this context, we are interested in the transcriptome de Bruijn graph (T-DBG), which is the De Bruijn graph constructed from *k*-mers present in the transcriptome (Fig. 1a) together with a path covering of the graph, where the paths correspond to transcripts (see Fig. 1b). This path covering of a T-DBG induces multi-sets on the vertices, called *k*-compatibility classes. A compatibility class can be associated to an error-free read by representing it as a path in the graph and defining the *k*-compatibility class of a path in the graph to be the intersection of the *k*-compatibility classes of its constituent *k*-mers (Fig. 1c). A key point is that the *k*-compatibility class of a read coincides with the equivalence class for large *k* (see Methods).

Previously, the equivalence classes of reads have been determined via the time-consuming alignment of the reads to the transcriptome. However since a hash of *k*-mers provides a fast way to determine their *k*-compatibility classes, the equivalence class of (error-free) reads can be efficiently determined by selecting suitably large *k* and then intersecting their constituent *k*-compatibility classes. The difficulty of implementing such an approach for RNA-Seq lies in the fact that reads have errors. However with very high probability, an error in a *k*-mer will result in it not appearing in the transcriptome and such *k*-mers are simply ignored. The issue of errors is also ameliorated by a technique that we implemented to improve the efficiency of pseudoalignment that removes redundant *k*-mers from the computation based on information contained in the T-DBG (see Methods). Because fewer *k*-mers are inspected, there is less opportunity for erroneous *k*-mers to produce misleading results. With pseudoalignments efficiently computable, we explored the use of the expectation-maximization (EM) algorithm applied to equivalence classes for quantification[5] (see Online Methods). Although the likelihood function is simpler than some other models used for RNA-Seq[2,3,9], its use has the advantage that the EM algorithm can be applied for many rounds very rapidly.

To validate and benchmark kallisto we first tested it on a set of 20 RNA-Seq simulations generated with the program RSEM[9]. The transcript abundances and error profiles for the simulated data were based on the quantification of sample NA12716_7 from the GEUAVDIS dataset[10], and to accord with GEUVADIS samples the simulations consisted of 30 million reads. We began by examining the quality of the kallisto pseudoalignments as compared to pseudoalignments extracted from Bowtie2 alignments. The two methods agreed exactly on the set of reported transcripts for 92.5% of the reads, but when they differed on the (pseudo)alignment of a read, Bowtie2 reported 5.18 transcripts on average compared to 3.58 for kallisto. Despite being much more specific than Bowtie2, kallisto had almost perfect sensitivity. The transcript of origin was contained in the set of reported transcripts for 99.56% of the reads, only 0.3% less than Bowtie2 (99.88%). On the real data used as the basis for the simulations (NA12716_7) the programs displayed similar characteristics. The two methods agreed exactly for 91.4% of reads where both (pseudo)aligned and for differing reads Bowtie2 aligned to 5.81 transcripts on average, versus 3.4 for kallisto. As expected, the number of (pseudo)aligned reads were lower with 82.8% of the reads aligned by Bowtie2 versus 88.0% pseudoaligned by kallisto.

Given the concordance of the kallisto and Bowtie (pseudo)alignments it is not surprising that we found the accuracy of kallisto to be comparable to that of widely used RNA-Seq quantification tools (Fig. 2a, Supp. Fig. 2), albeit with substantial improvement over Cufflinks[2] and Sailfish[5]. The inferior performance of Cufflinks can be attributed to its limited application of the EM algorithm in cases where reads multi-map across genomic locations[11]. And unlike Sailfish, which shreds reads into *k*-mers for fast hashing resulting in a loss of information, kallisto's pseudoalignments explicitly preserve the information provided by *k*-mers across reads (Supp. Fig. 1). We also examined the performance of kallisto on paralogs since they are particularly difficult to quantify. All programs have reduced performance due to the similarity among genes within a family but kallisto remains highly competitive, again almost matching RSEM's performance (Supp. Fig. 3a,b). To test kallisto's suitability for allele specific expression (ASE) quantification, we also simulated reads from a transcriptome with two distinct haplotypes. The median percent error for kallisto was 12.2% vs. 11.3% for RSEM, 14.8% for eXpress and 14.6% for Sailfish, showing that kallisto is suitable for ASE (it is important to note again, that the simulation was based on RSEM, both for generating the parameters and then the data).

The most striking aspect of kallisto is its speed: each simulation was processed on average in less than 3 minutes on a single core, so fast that when quantifying all 20 simulations on a multi-core server the program was I/O bound. Even so, the total runtime for kallisto on the simulations was 5.29 minutes (Fig. 2b). A simple word count of a simulated dataset took 43 seconds, providing a lower bound for optimal quantification time and revealing kallisto to be near-optimal in speed. The software is also memory efficient, requiring a maximum of 3.2 Gb of RAM per sample. This allowed us to test kallisto on a laptop where the 30 million read simulations were each processed in less than 5 minutes on a 2014 MacBook Air, demonstrating that with kallisto RNA-Seq analysis of even large datasets is tractable on non-specialized hardware.

The speed of kallisto is not just a matter of convenience; it also allows for quantifying the uncertainty of abundance estimates via the bootstrap technique of repeating analyses after resampling with replacement from the data. Here, after the equivalence classes of the original reads have been computed, that requires merely sampling multinomially from the equivalence classes according to their counts and then running the EM algorithm on those newly sampled equivalence class counts. We explored the accuracy with which the bootstrap can estimate the uncertainty inherent in a dataset by examining repeated 30 million read subsamples of a deep 216 million read human RNA-Seq dataset from the SEQC-MAQCIII[12] consortium (Fig. 3). The bootstrap was performed on only a single sample of 30 million reads, yet the variance in estimates correlates highly (R=0.957) with the variance of abundance estimates obtained from the other subsamples. While it is expected that the variance on abundance estimates should increase approximately linearly with abundance[13], our results show that there is high variability in uncertainty of estimates as a result of the complex structure of similarity among transcripts, especially multiple isoforms of genes. A naïve attribution of “Poisson variance” to the “shot-noise” in read counts from transcripts, as commonly utilized currently in RNA-Seq, is revealed to be a poor proxy for the true variance (Supp. Fig. 4). Thus, the bootstrap should prove to be invaluable in downstream applications of RNA-Seq, as kallisto now allows for the uncertainty in estimates to be factored in to downstream statistical computations.

The simplicity of kallisto means that the software has few parameters; only the *k*-mer length and the mean of the fragment length distribution are required for quantification. The latter is estimated during run-time when paired-end reads are provided. The *k*-mer length must be large enough that random sequences of length *k* do not match to the transcriptome, and short enough to ensure robustness to errors. Subject to those constraints the performance of kallisto is robust to the *k*-mer length chosen (Supp Fig. 5). Although we have focused on the performance of kallisto on RNA-Seq, the method should be generally applicable to quantification of sequence census datasets[14].

# Online Methods

## Index construction

The construction of the index starts with the formation of a colored de Bruijn graph[15] from the transcriptome, where the colors correspond to transcripts. In the colored transcriptome de Bruijn graph each node corresponds to a *k*-mer and every *k*-mer receives a color for each transcript it occurs in. Contigs are formed from linear stretches of the de Bruijn graph that have identical colorings. This ensures that all *k*-mers in a contig are associated with the same equivalence class (the converse is not true: two different contigs can be associated with the same equivalence class). Once the graph and contigs have been constructed, kallisto stores a hash table mapping each *k*-mer to the contig it is contained in, along with the position within the contig. This structure is called the “kallisto index”.

For error-free reads there can be a difference between the equivalence class of a read and the intersection of its *k*-compatibility classes. But for a read of length *l* this can only happen if there are two transcripts that have the same *l-k+1* *k*-mers occuring in different order. This is unlikely to happen for large *k* because it would imply that the T-DBG has a directed cycle shorter than *l-k+1*. This fact also provides a criterion that can be tested.

## Pseudoalignment

Reads are pseudoaligned by looking up the *k*-compatibility class for each *k*-mer in the read with the kallisto index, and then intersecting the identified *k*-compatibility classes. In the case of paired-end reads the *k*-compatibility class lookup is done for both ends of the fragment and all the resulting classes are intersected. To speed up the processing, kallisto uses the positional information stored in the index: each time a *k*-mer is looked up using the hash, kallisto finds the distances to the junctions at the end of its contig in the T-DBG. Since all *k*-mers in a contig of the T-DBG have the same *k*-compatibility class, they will not affect the intersection and therefore looking them up in the hash provides no new information. This observation is leveraged in kallisto, which skips over redundant *k*-mers in the read the maximum possible distance, i.e. the minimum of the junction distance or the distance to the end of the read. To ensure that the skip is consistent with the T-DBG, kallisto checks the last *k*-mer of the skip to ensure the *k*-compatibility class is equal as expected. In rare case when there is a mismatch, kallisto defaults to examining each *k*-mer of the read. For the majority of reads, kallisto ends up performing a hash lookup for only two *k*-mers (Supp. Fig. 6).

## Quantification

In order to rapidly quantify transcript abundances from pseudoalignments, kallisto makes use of the following form of the likelihood function for RNA-Seq:

$$L(\alpha) \propto \prod_{f \in F} \sum_{t \in T} y_{f,t} \frac{\alpha_t}{l_t} = \prod_{e \in E} \left( \sum_{t \in e} \frac{\alpha_t}{l_t} \right)^{c_e}.$$

In the likelihood function *F* is the set of fragments, *T*, the set of transcripts, $l_t$ is the (effective length[3]) of transcript *t* and $y_{f,t}$ is a compatibility matrix defined as 1 if fragment *f* is compatible with *t* and 0 otherwise. The parameters are the $\alpha_t$, the probabilities of selecting fragments from transcripts. The likelihood can be rewritten as a product over equivalence classes, in which similar summation terms have been factored together. In the factorization the numbers $c_e$ are the number of counts observed from equivalence class *e*. When the likelihood function is written in terms of the equivalence classes, the equivalence class counts are sufficient statistics and thus in the computations are based on a much smaller set of data (usually hundreds of thousands of equivalence classes instead of tens of millions of reads). The likelihood function is iteratively optimized with the EM algorithm with iterations terminating when for every transcript *t*, $\alpha_t N > 0.01$ (*N* is the total number of fragments) changes less than 1% from iteration-to-iteration.

The transcript abundances are output by kallisto in transcripts per million[9] (TPM) units.

## Bootstrap

The bootstrap is highly efficient in kallisto both because the EM algorithm is very fast and because the sufficient statistics of the model are the equivalence class counts. This latter fact means that bootstrap samples can be very rapidly generated once pseudoalignment of the fragments is completed. With the *N* original fragments having been categorized by equivalence class, generating a new bootstrap sample consists of sampling *N* counts from a multinomial distribution over the equivalence classes with the probability of each class being proportional to its count in the original data. The transcript abundances for these new samples are then recomputed using the EM algorithm.

In kallisto the number of bootstraps to be performed is an option passed to the program, and due to the fact that a large amount of data can be produced the output is compressed in HDF5. The HDF5 files can be read into another program for processing (e.g. R) or can be converted to plaintext using kallisto.

## Simulations and analysis

The parameters and procedures used for the results and figures in the paper are available via a snakefile[16] at https://github.com/pachterlab/kallisto_paper_analysis.

## Software

The kallisto program is available for download from http://pachterlab.github.io/kallisto.

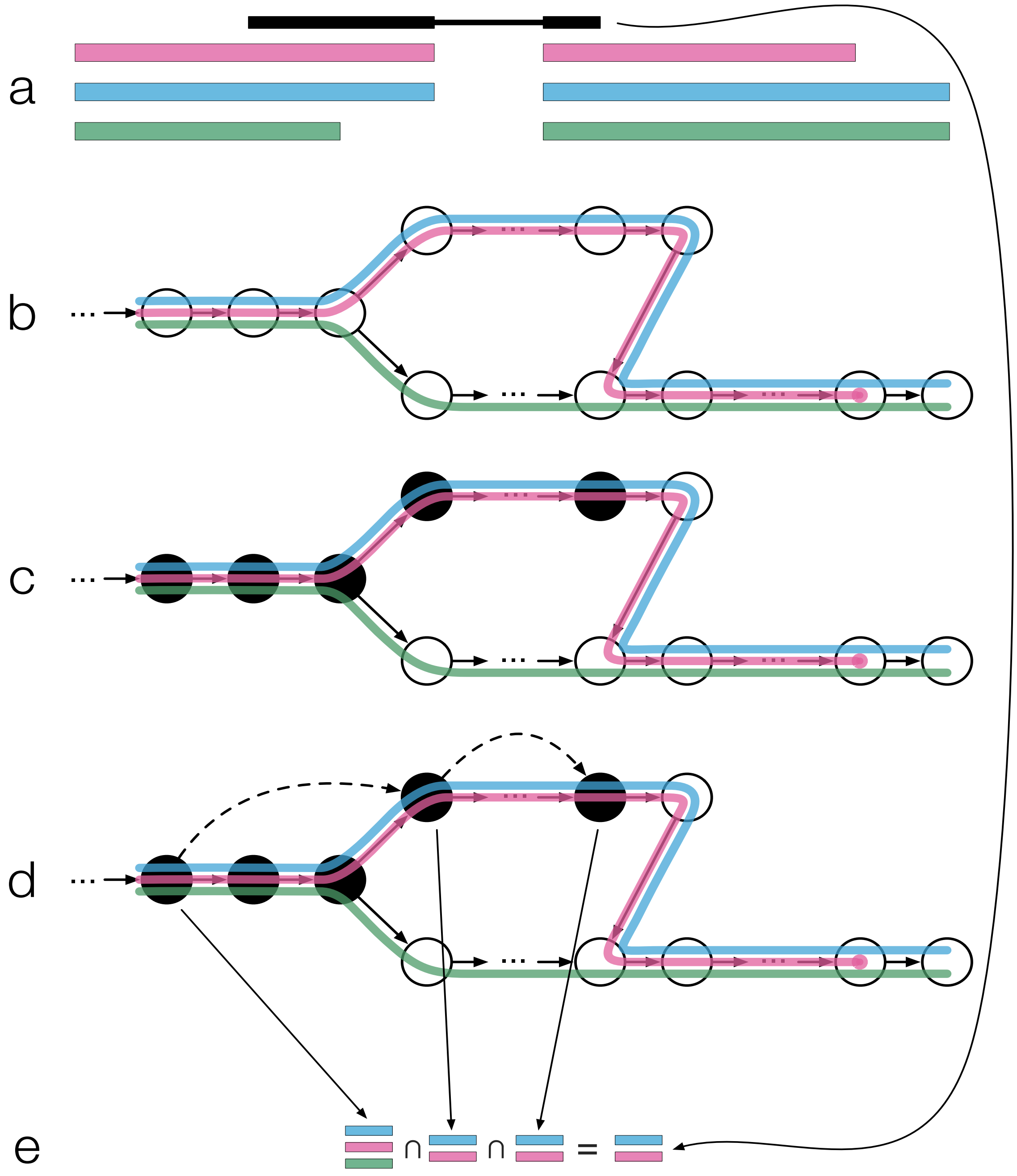

Figure 1: Overview of kallisto. (a) The input consists of a reference transcriptome and reads from an RNA-Seq experiment. (b) An index is constructed by creating the Transcriptome de Bruijn Graph (T-DBG) where nodes are *k*-mers, each transcript corresponds to a path and the path cover of the transcriptome induces a *k*-compatibility class for each *k*-mer. (c) Conceptually, the *k*-mers of a read are hashed (black nodes) to find the *k*-compatibility class of a read. (d) Skipping uses the information stored in the T-DBG to skip *k*-mers that are redundant due to having the same *k*-compatibility class. (e) The *k*-compatibility class of the read is determined by taking the intersection of the *k*-compatibility classes of its constituent *k*-mers.

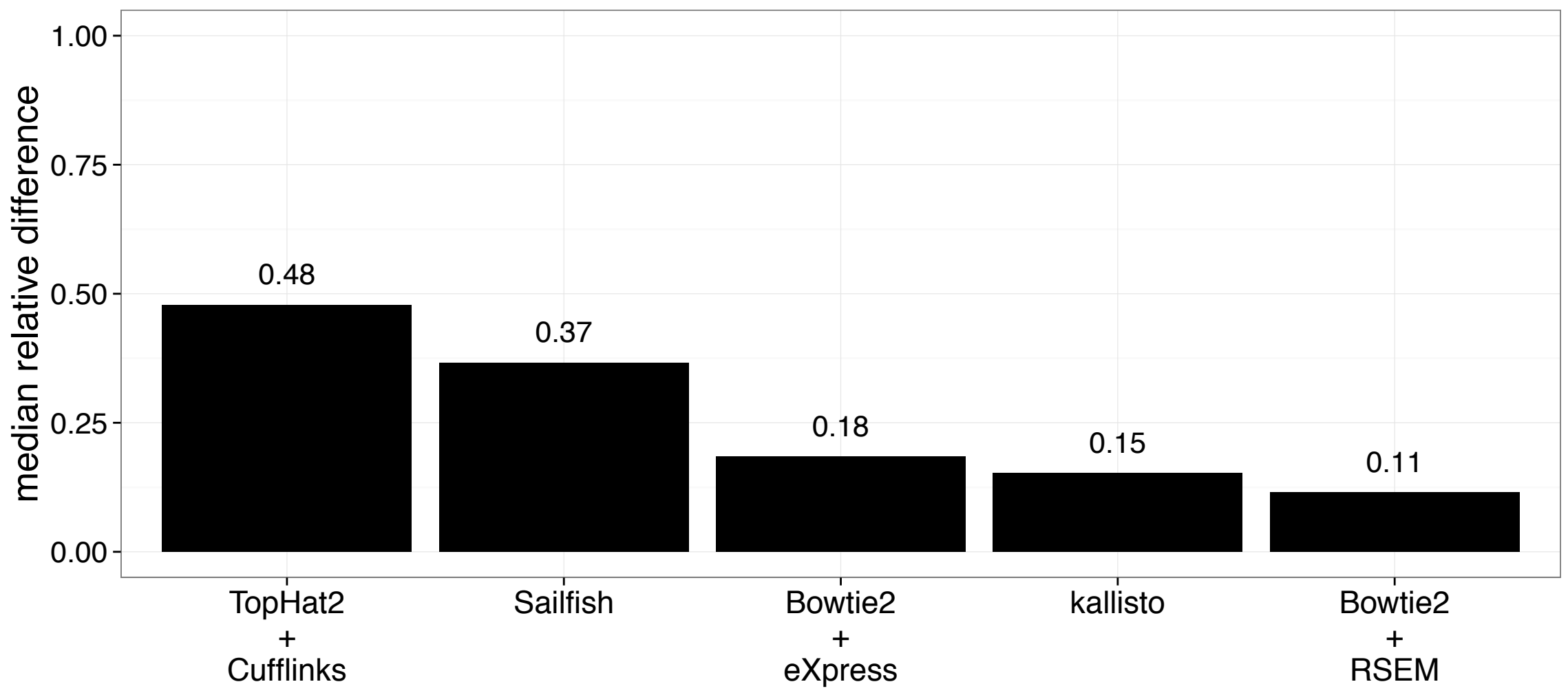

Figure 2a: Accuracy of kallisto, Cufflinks, Sailfish, eXpress and RSEM on 20 RSEM simulations of 30 million 75bp paired-end reads based on the abundances and error profile of Geuvadis sample NA12716 (selected for its depth of sequencing). For each simulation we report the accuracy as the median relative difference in the estimated read count of each transcript. Estimated counts were used to separate between the assignment of ambiguous reads and the estimation of effective lengths of transcripts. The values reported are means across the 20 simulations (the variance was too small for this plot). Relative difference is defined as the absolute difference between the estimated abundance and the ground truth divided by the average of the two.

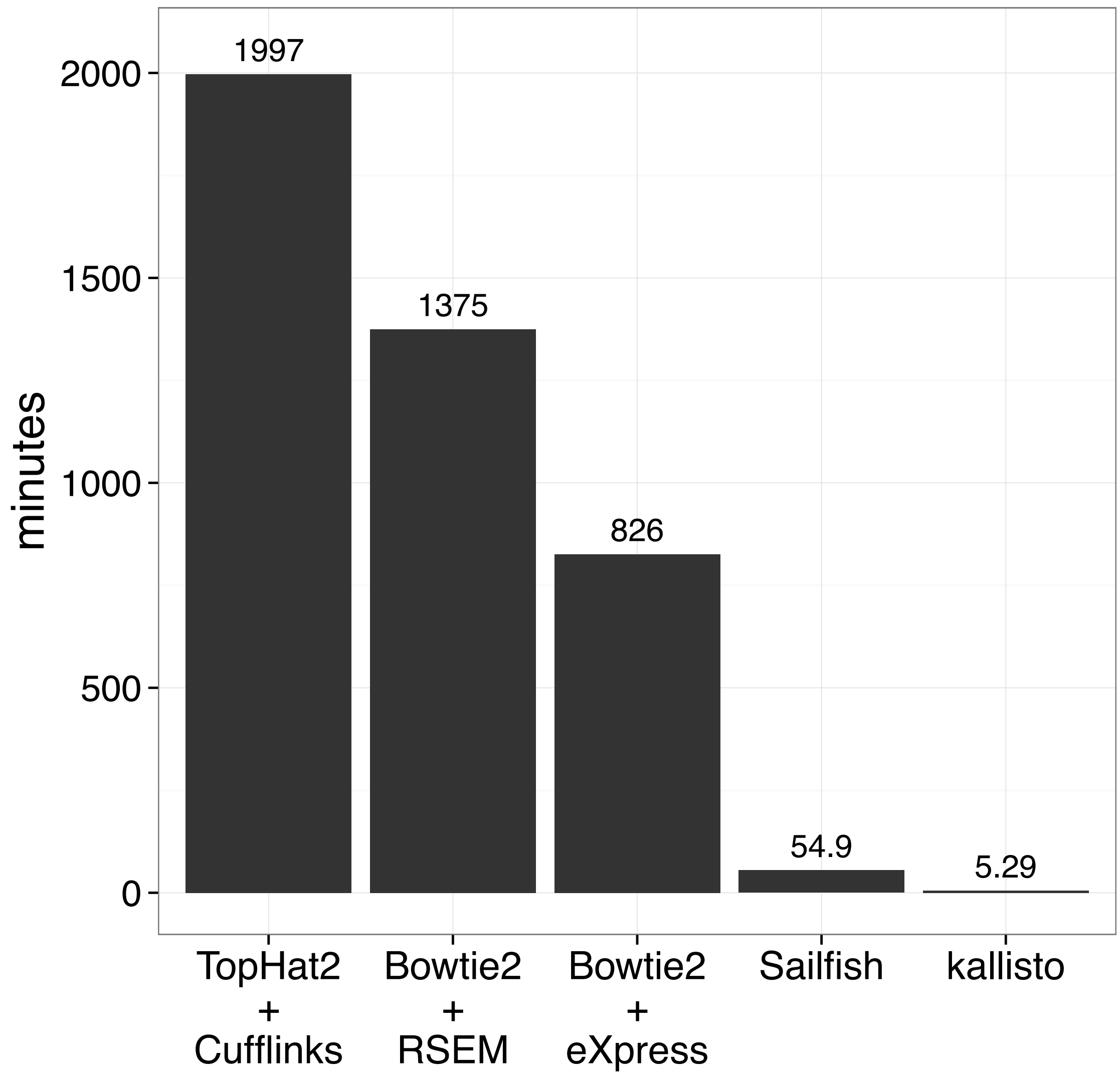

Figure 2b: Total running time in minutes for processing the 20 simulated datasets of 30 million paired-end reads described in Figure 2a. All processing was done using 20 cores with programs being run with 20 threads when possible (Bowtie2, TopHat2, RSEM, Cufflinks) and 20 parallel processes when not (eXpress, kallisto).

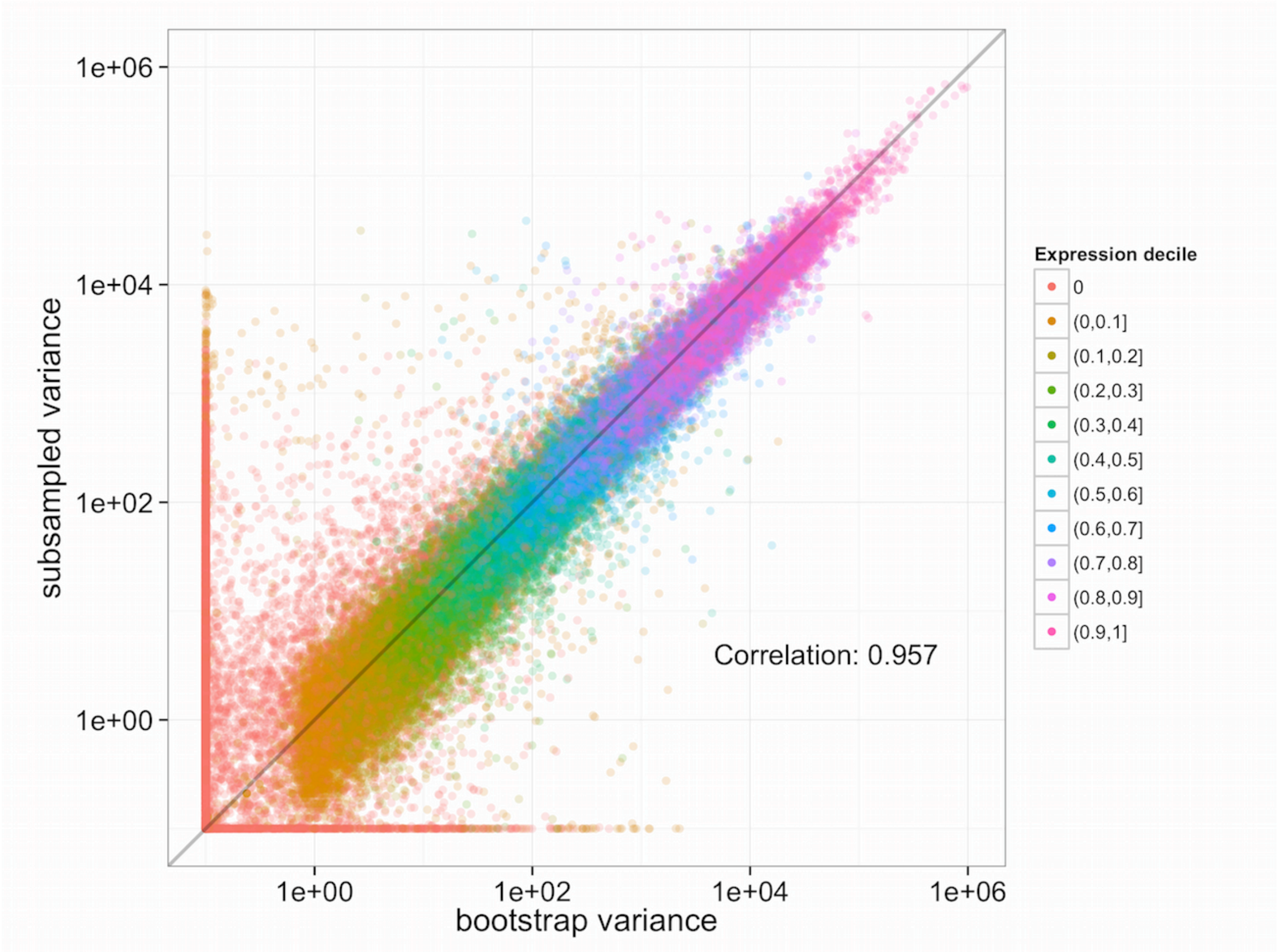

Figure 3: Comparison of technical variance in abundances. The data comes from a single library with 216M, 101bp paired-end reads sequenced. Each point corresponds to a transcript and is colored by the decile of its expression level in the single bootstrapped subsample. The Y-axis represents variance of abundance estimates across 40 subsamples, with 30M reads in each subsample. The X-axis represents variance as computed from 40 bootstraps of a single subsampled dataset of 30M reads. The red lines emanating from the lower left corner consist of transcripts that have an estimated abundance of zero in the single bootstrapped experiment, but show expression in some of the subsamples (12968 transcripts), and vice versa (720 transcripts).

# Supplementary figures

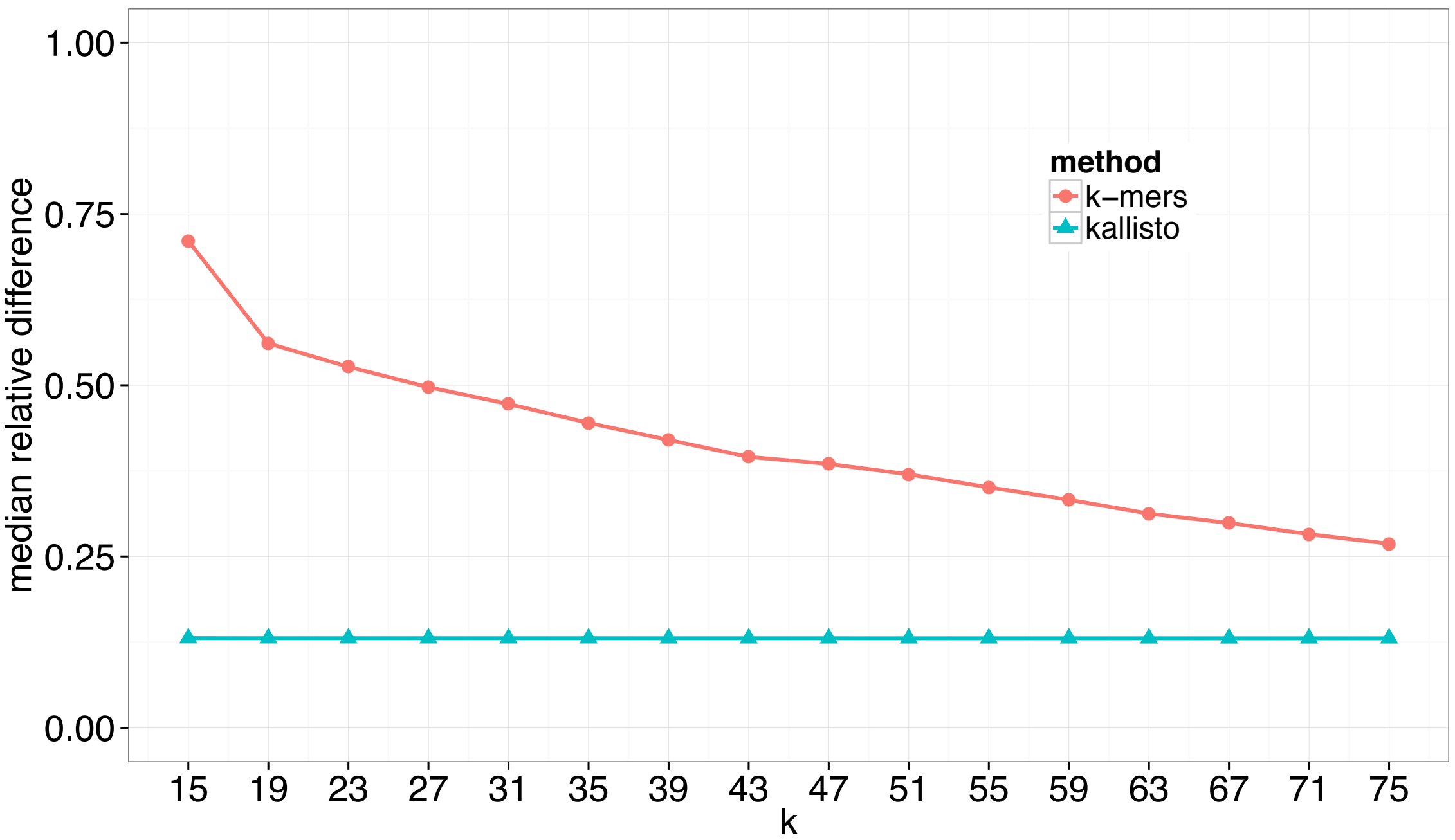

Supplementary Figure 1: Median relative difference for abundance estimates using varying values of *k* on a dataset of 30 million 75bp paired-end reads that were simulated without errors. The “*k*-mers method” uses the *k*-compatibility of each *k*-mer independently and runs the EM algorithm on *k*-mers, whereas kallisto uses the intersection of *k*-compatibility classes across both ends of a read. Even for *k*=75, the full read length in the simulation, independent use of *k*-mers results in a significant drop in accuracy due to the loss of paired-end information.

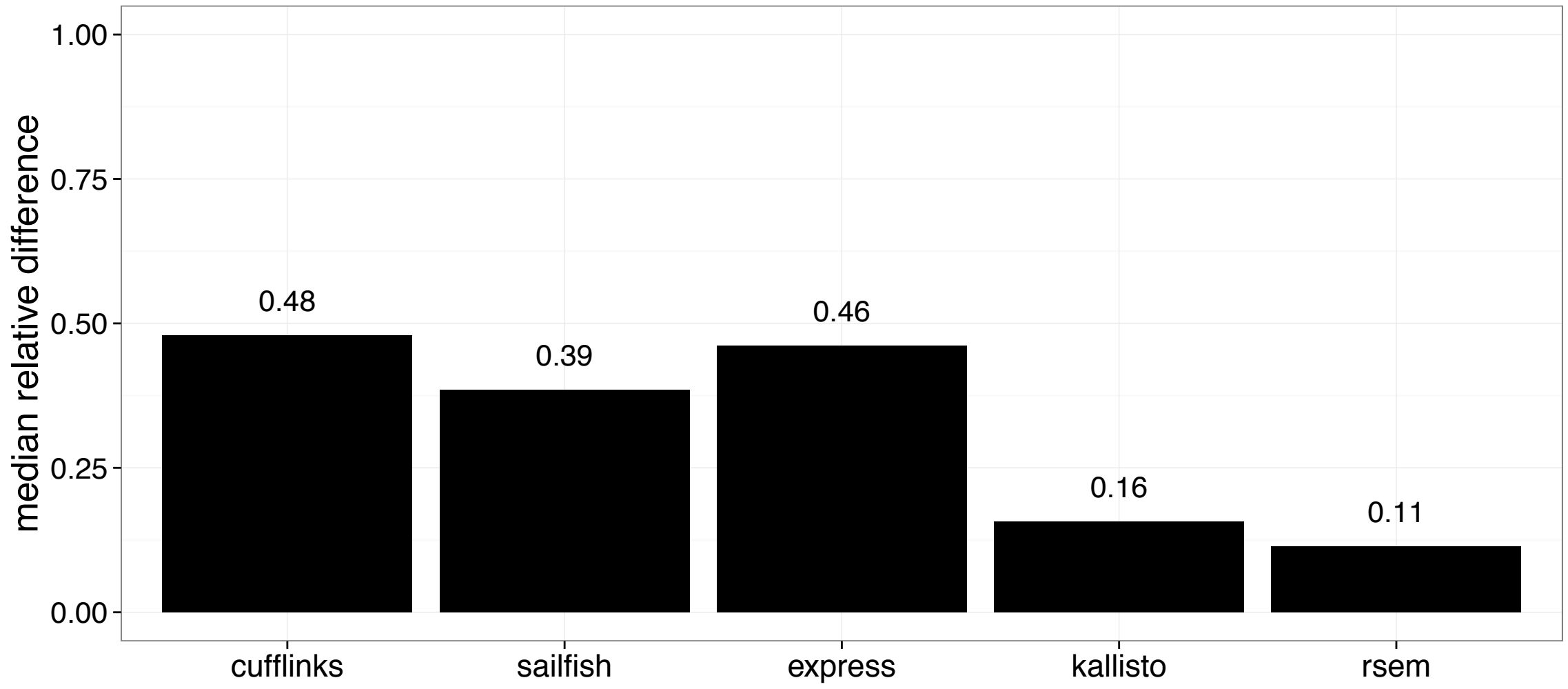

Supplementary Figure 2: Accuracy of kallisto, Cufflinks, Sailfish, eXpress and RSEM on 20 RSEM simulations of 30 million 75bp paired-end reads based on the TPM estimates and error profile of Geuvadis sample NA12716 (selected for its depth of sequencing). For each simulation we report the accuracy as the median relative difference in the estimated TPM value of each transcript. The values reported are means across the 20 simulations (the variance was too small for this plot). Relative difference is defined as the absolute difference between the estimated TPM values and the ground truth divided by the average of the two.

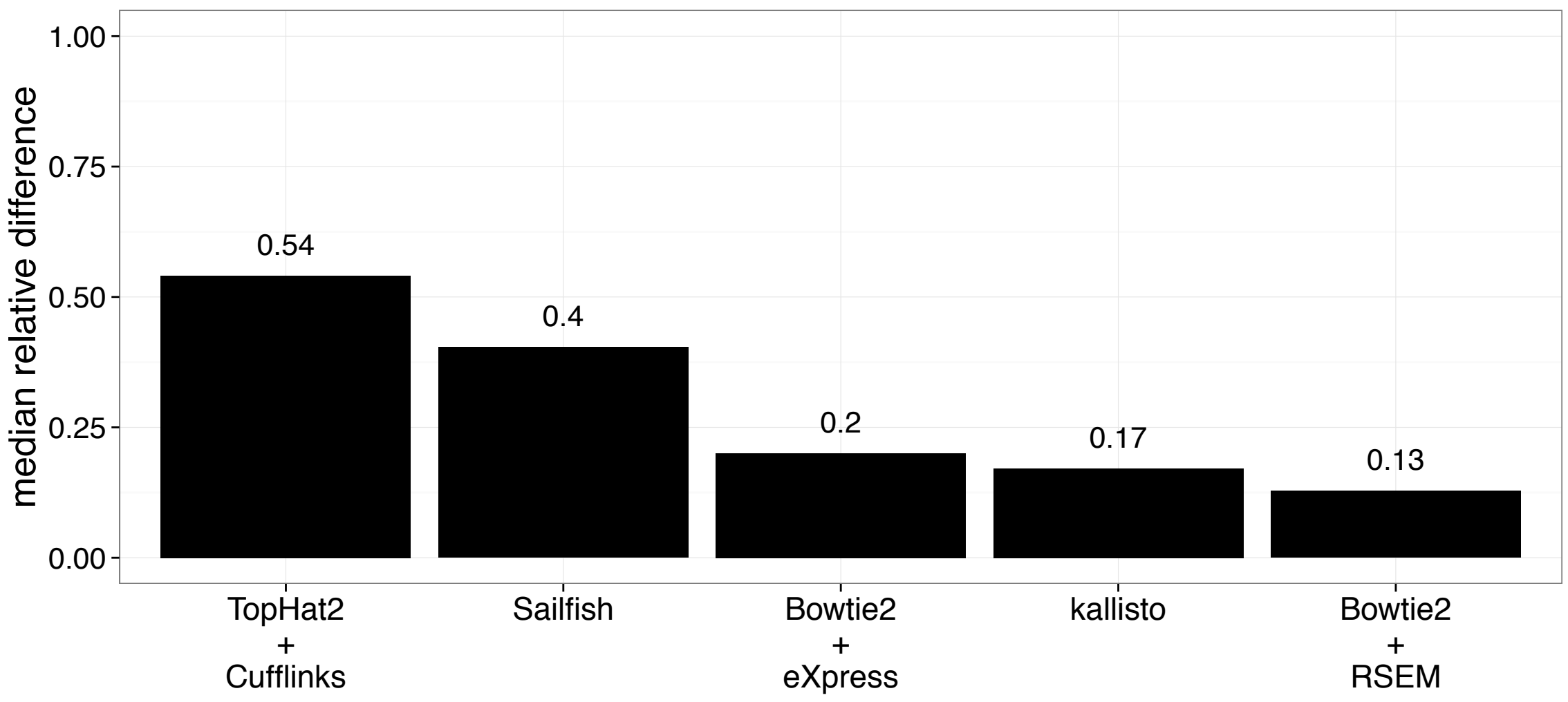

Supplementary Figure 3a: Performance of different quantification programs on the set of paralogs in the human genome supplied by the Duplicated Genes Database (http://dgd.genouest.org). This set includes 8,636 transcripts in 3,163 genes.

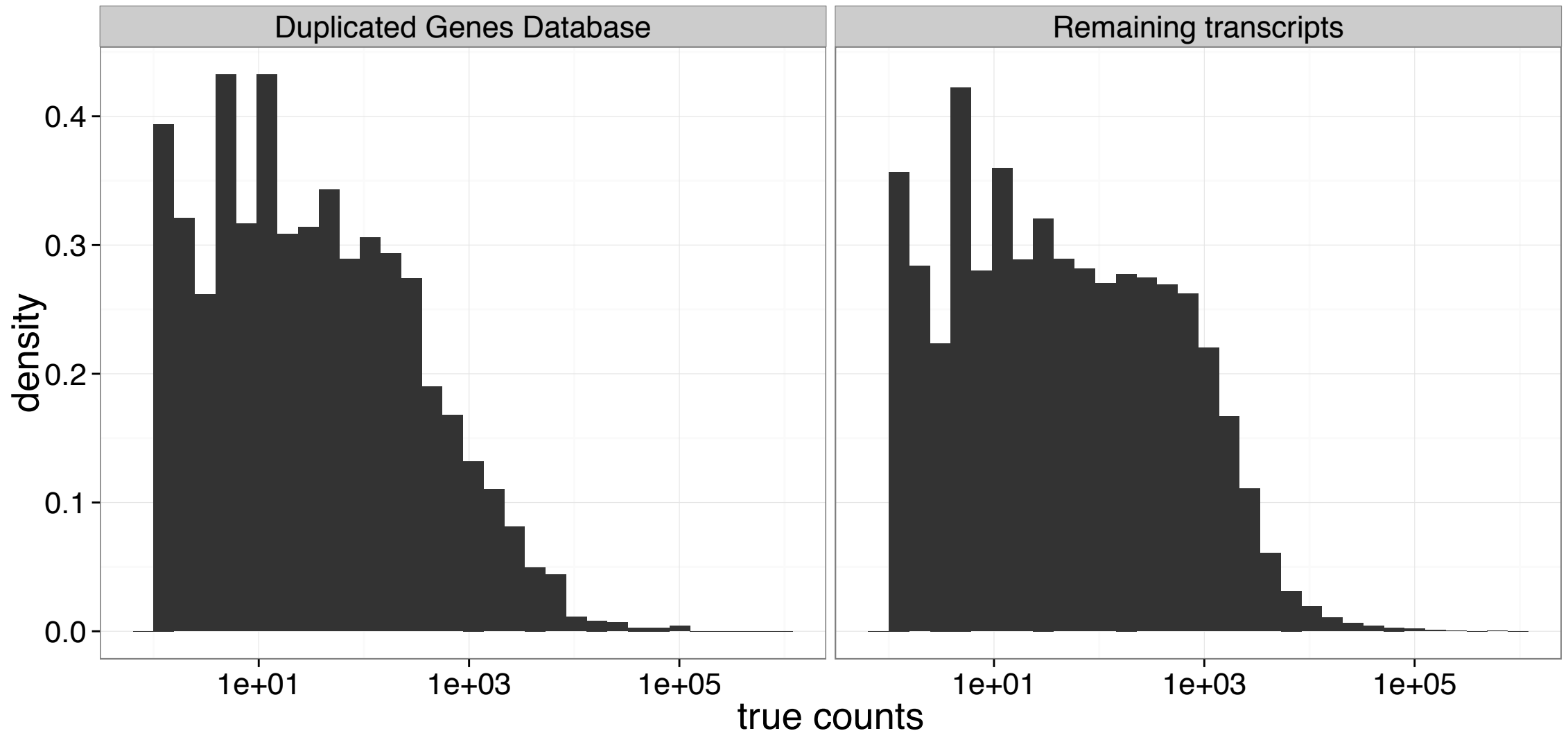

Supplementary Figure 3b: Count distribution of one simulation. The left panel contains the transcripts used in Supplementary Figure 2. The right panel contains the remaining transcripts. The x-axis is on the log scale. Both distributions appear very similar, suggesting that the drop in performance in Supplementary Figure 2a is from sequence similarity and not oddities in the distribution such as very low counts.

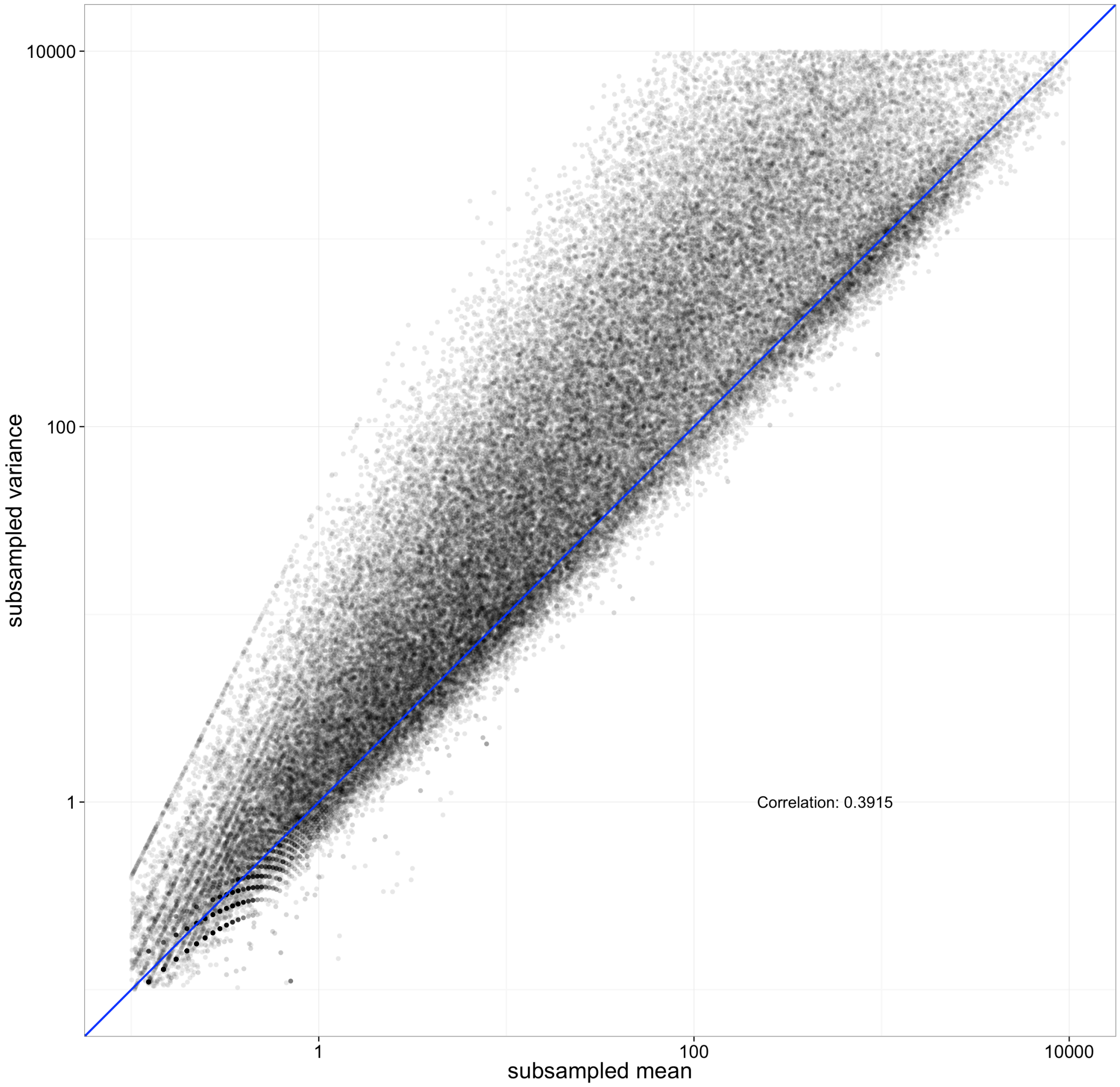

Supplementary Figure 4a: Relationship between the mean and variance of estimated counts for each transcript (x and y axes are on log scale) based on 40 subsamples of 30M reads from a dataset of 216M PE reads. The *x*-axis is the mean of each count estimate calculated across the subsamples. The *y*-axis is the variance of the count estimates calculated across subsamples.

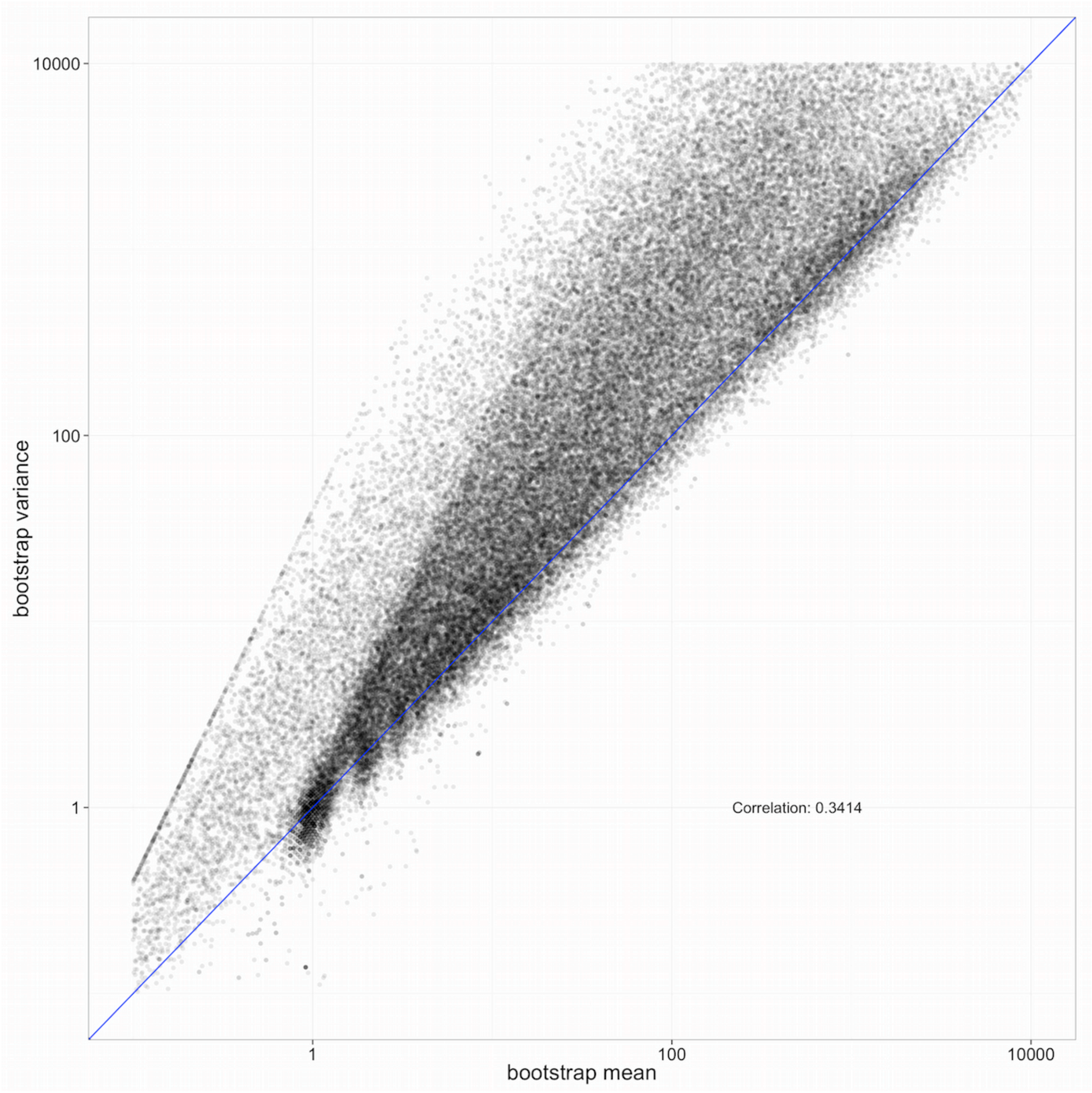

Supplementary Figure 4b: Relationship between the mean and variance of estimated counts for each transcript (x and y axes are on log scale) based on 40 bootstraps of a single subsample of 30M reads from the same 216M PE read dataset. The *x*-axis is the mean of the count estimates calculated across the 40 bootstraps. The *y*-axis is the variance of the count estimates calculated across the 40 bootstraps.

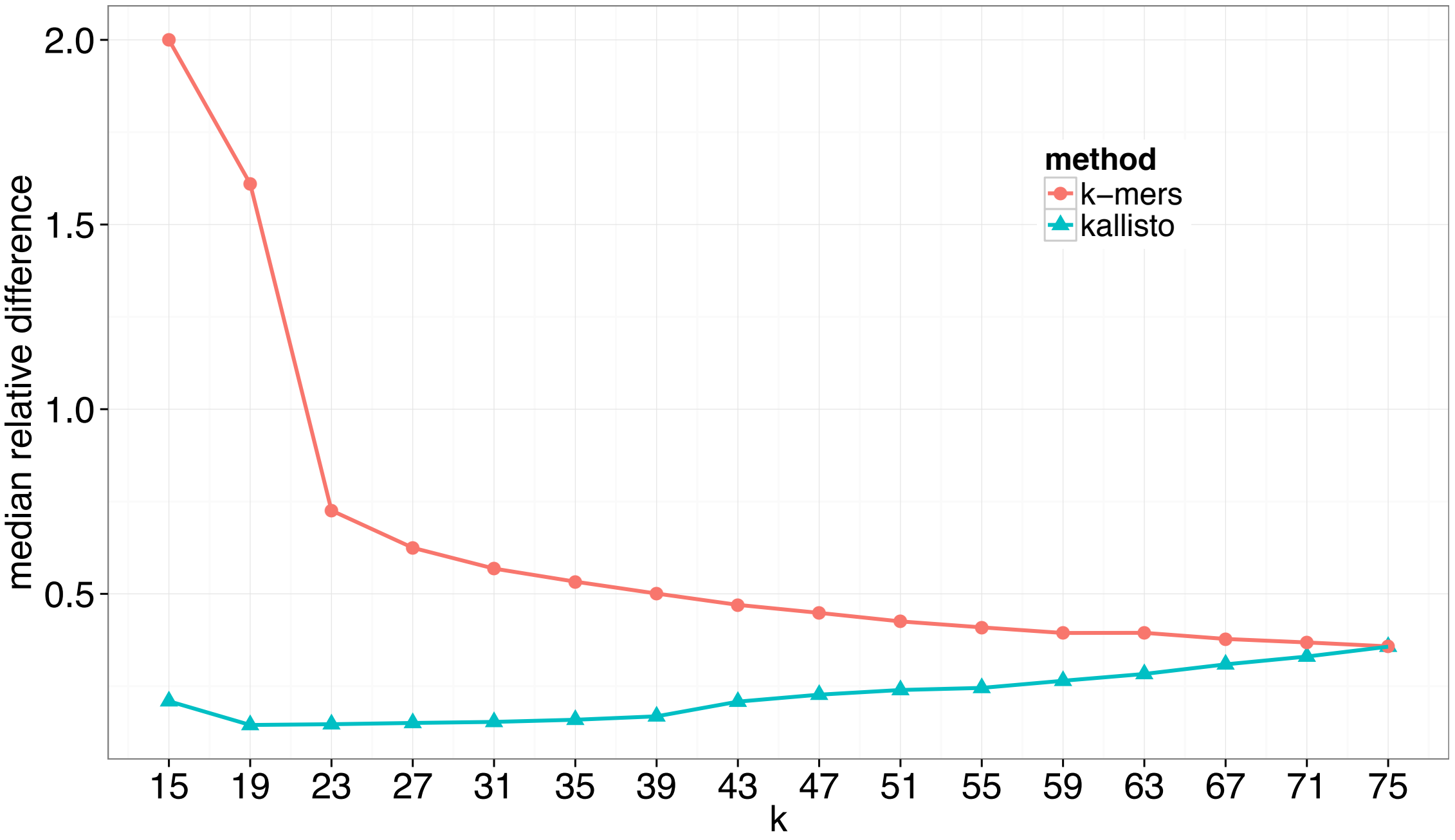

Supplementary Figure 5: Median relative difference from 30M 75bp PE reads simulated with error for different values of *k*. The "*k*-mers method" uses the *k*-compatibility of each *k*-mer independently and runs the EM algorithm on *k*-mers, whereas kallisto uses the intersection of *k*-compatibility classes across both ends of reads. When there are errors in the reads, kallisto requires smaller *k*-mer lengths for robustness in pseudoalignment.

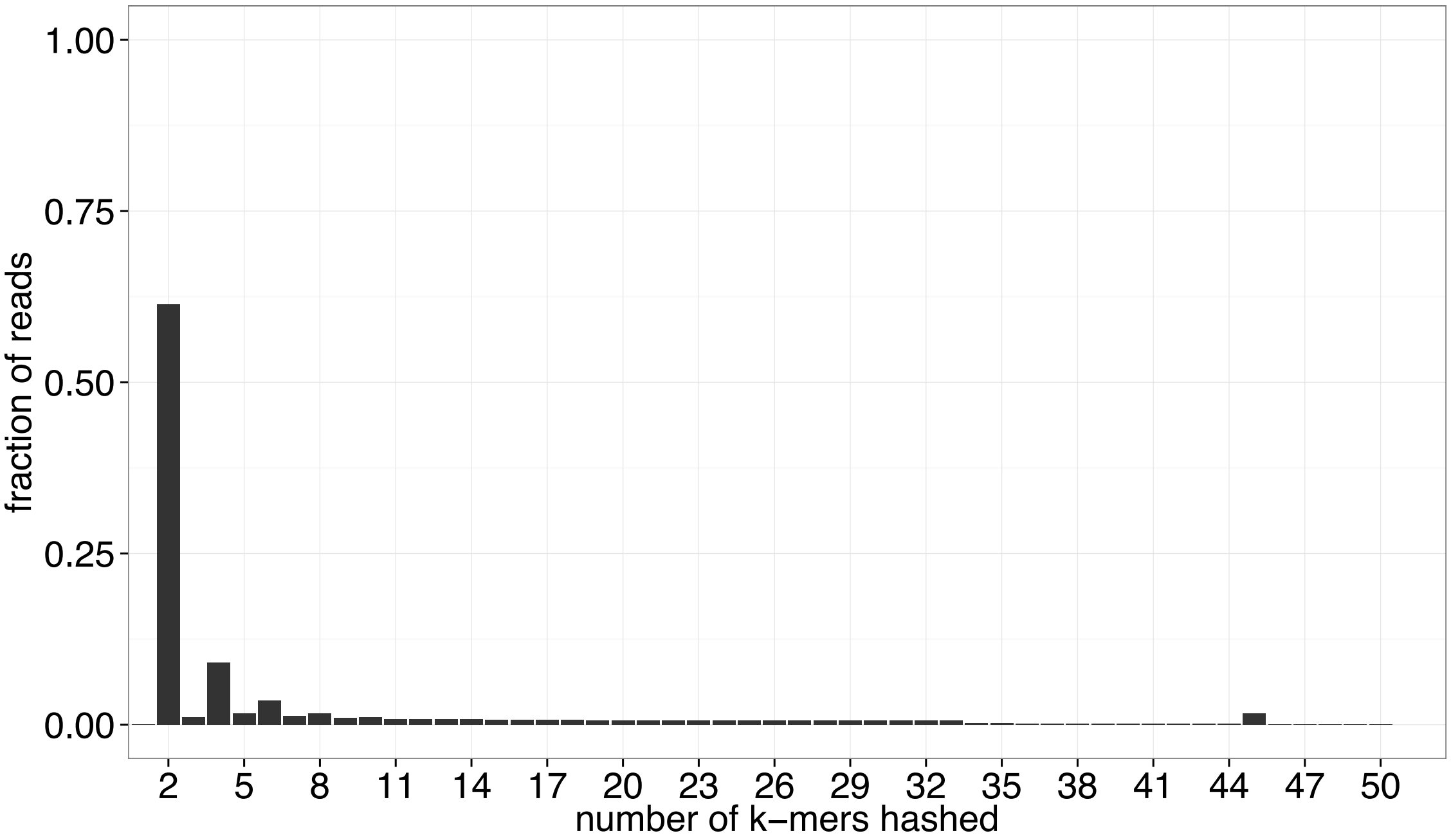

Supplementary Figure 6: The distribution of the number of *k*-mers hashed per read for k=31. Note that for the majority of reads (61.35%) only two *k*-mers are hashed. This happens when the entire read pseudoaligns to a single contig of the T-DBG and we can skip to the end of the read. Since we also check the last *k*-mer we can skip over, the most common cases are checking 2, 4, 6, and 8 *k*-mers. Only 1.6% of reads required hashing every *k*-mer of the read.